# Scaling of variations in traveling distances and times of taxi routes

Xiaoyan Feng[a], Huijun Sun[a,*], Bnaya Gross[b], Jianjun Wu[c,*], Daqing Li[d,e], Xin Yang[c,*], Dong Zhou[d], Ziyou Gao[a] and Shlomo Havlin[b,*]

[a]School of Traffic and Transportation, Beijing Jiaotong University, Beijing 100044, China;

[b]Department of Physics, Bar-Ilan University, Ramat-Gan 52900, Israel;

[c]State Key Laboratory of Rail Traffic Control and Safety, Beijing Jiaotong University, Beijing 100044, China;

[d]School of Reliability and Systems Engineering, Beihang University, Beijing 100191, China;

[e]National Key Laboratory of Science and Technology on Reliability and Environmental Engineering, Beijing 100191, China

**Abstract.** The importance of understanding human mobility patterns has led many studies to examine their spatial-temporal scaling laws. These studies mainly reveal that human travel can be highly non-homogeneous with power-law scaling distributions of distances and times. However, investigating and quantifying the extent of variability in time and space when traveling the *same* air distance has not been addressed so far. Using taxi data from five large cities, we focus on several novel measures of distance and time to explore the spatio-temporal *variations* of taxi travel routes relative to their *typical* routes during peak and nonpeak periods. To compare all trips using a single measure, we calculate the distributions of the *ratios* between actual travel distances and the average travel distance as well as between actual travel times and the average travel time for all origin destinations (OD) during peak and nonpeak periods. In this way, we measure the scaling of the distribution of all single trip paths with respect to their mean trip path. Our results surprisingly demonstrate very broad distributions for both the distance ratio and time ratio, characterized by a long-tail power-law distribution. Moreover, all analyzed cities have larger exponents in peak hours than in nonpeak hours. We suggest that the interesting results of shorter trip lengths and times, characterized by larger exponents during rush hours, are due to the higher availability of travelers in rush hours compared to non-rush hours. We also find a high correlation between distances and times, and the correlation is lower during peak hours than during nonpeak hours. The reduced correlations can be understood as follows. Due to the high availability of passengers in peak periods more drivers choose long distances to save time compared to nonpeak periods. Furthermore, we employed an indeterminate traffic assignment model, which supports our finding of the power-law distribution of the distance ratio and time ratio for human mobility. Our results can help to assess traffic conditions within cities and provide guidance for urban traffic management.

**Keywords:** human mobility; route variability; scaling laws; spatiotemporal; correlation

* Correspondence and requests for materials should be addressed to H.J.S. (email: hjsun1@bjtu.edu.cn) or to J.J.W. (email: jjwu1@bjtu.edu.cn) or to X.Y. (email: xiny@bjtu.edu.cn) or to S.H. (email: havlins@gmail.com)

# Introduction

Studying human movement behavior has been regarded as a long-standing fundamental and challenging task. Understanding human mobility patterns is of much importance in many aspects, such as urban planning [1,2], traffic engineering [3,4], epidemic spreading [5–7], and emergency management [8,9]. Initially, researchers relied on using and analyzing human activity data collected from travel surveys or observations [10,11]. With the widespread use of mobile positioning technologies in people's daily lives, massive individual mobility data becomes available, including GPS trajectories of vehicles [12,13] and humans [14,15], cell phone records [16–18], and check-ins of online social network accounts [19,20]. Such big data offers an excellent opportunity to uncover human mobility patterns more accurately and understand their underlying mechanisms more deeply.

In the last decade, human mobility patterns on different geographical scales have been extensively studied. In large scale of space, including trips between countries or cities, many studies have found that statistical patterns of human movements exhibit a long-tail Lévy walk characteristics [21–24]. For example, the aggregated trip lengths and waiting time distributions characterizing human trajectories have been found to be fat-tailed power laws through investigating the dispersal of bank notes [21] and mobile phone records [22]. In order to understand the observed scaling laws, several microscopic models have been developed for the movement process of individuals, capturing some dynamic features [21,23,25,26]. Furthermore, a number of macroscopic models have been developed to predict the mobility flow between spatial locations, and these models considered the interactions among individuals [27–32]. Short-scale mobility within the range of a city has attracted particular attention from researchers, as cities are concentrated areas of human activities, and intra-city movement is a significant part of citizens' lives. However, unlike the mobility patterns observed at large spatial scales, human movement within cities tends to exhibit different scaling behaviors. Jiang et al. [33] found that the distance traveled by cab passengers follows a two-phase power-law distribution. Yao and Lin [13] analyzed taxi trajectories in a South China city and found a power-law behavior of travel distances. In contrast, various other studies on datasets of taxis [34–36], private cars [12], and mobile phones [17] show exponential distributions of travel distances or displacements. Similarly, studies of the traveling time have found different distributions such as exponential distribution [35] and lognormal distribution [34]. Several works have been based on simulation models to reproduce the observed distributions [13,19,33] and explain them from different perspectives, such as the place density [19], time and fare [13]. Currently, researchers have richly explored the human mobility patterns, revealing that human movements have a very broad range of scales in terms of time and distance. However, the extent of variations with respect to the *typical* (average) distance and time for the same OD pair, to the best of our knowledge, has not been addressed so far. For example, we ask how many trips deviate from the typical travel path of a given OD and how much do they deviate?

In this paper, we explore the scale of deviations between single trip paths and their typical (average) path by measuring the distributions of the distance *ratio* and the time *ratio* between a single trip and the average trip for all OD pairs. For each OD, we evaluate the average distance and time of all taxis and analyze the distribution of the above ratios for all ODs. Based on high-resolution taxi data from five cities, we surprisingly find, in all analyzed cities, scaling characterized by long-tail power-law distributions for both distance and time ratios and compare the scaling during peak and nonpeak periods. Interestingly, we find that in rush hours the broadness of the variations is narrower compared to non-rush hours. We explain these shorter relative distances and times by the availability of significantly more passengers in rush hours, thus motivating the drivers to make shorter trips in rush hours. Additionally, based on an indeterminate traffic assignment model [37], we support the scaling laws of these two ratios. Our findings suggest the existence of intrinsic behavior behind taxi trips, resulting from drivers' individual choices and influenced by drivers' estimated travel costs (primarily travel time). The scaling laws found here could potentially help to understand urban traffic conditions and to develop appropriate traffic management methods.

**Results**

Our study uses taxi datasets from three major cities (Beijing, Chengdu, and Shenzhen) in China and two major cities (New York and Chicago) in the United States (details are in SI, Table S1). For these five cities, only weekdays' data are studied. These datasets include 5,133,615 trips in Beijing during 10-weekdays, 7,216,951 trips in Chengdu during 14-weekdays, 4,002,107 trips in Shenzhen during 10-weekdays, 7,339,443 trips in New York during 20-weekdays, and 2,317,823 trips in Chicago during 40-weekdays. The data includes for each trip, the taxi id, pick-up timestamp, drop-off timestamp, pick-up location, drop-off location, travel distance, and travel time.

To study the diversity of travel routes, we divide the area of the four cities except for Chicago into square grids (Fig. 1A) with a side length of 0.5 km, and Chicago is divided by the official census area (SI, Fig. S1). A grid or census represents a traffic zone. Each taxi trip starts within its origin zone (O) and ends within its destination zone (D). Thus, each OD pair is assigned with many taxi trips during the day. In Fig. 1A we demonstrate the grid and the OD while in Fig. 1B we show two different travel routes between the same OD pair, including the typical path with a length close to the average distance and the detour path with a length much longer than the average distance. In this demonstration the distance traveled along the detour path is approximately twice the length of the typical (average) path, but the time they spent may be longer or shorter. Additionally, between this OD pair, there are trips that are even seven and more times longer than the average (Fig. 1C). Thus, we ask here how many deviated trips exist and how much they deviate in both distance and time in the whole network.

Considering the temporal variability of taxi travel [38], we distinguish and divide the taxi trips into two periods: peak hours and nonpeak hours. Peak hours vary between cities and are the period of the day with the highest traffic flow (around 7:00-9:00 and 17:00-19:00 in most cities), and nonpeak hours are all other times. Then, we demonstrate the scales of single trip paths deviating from the average travel path by analyzing the distribution of the distance ratio $r_d$ and the time ratio $r_t$. The distance ratio of a single trip, $r_d$, is defined as the actual travel distance of the trip divided by the average travel distance of all trips in this OD, which is calculated separately for peak hours and nonpeak hours. Thus, we obtain the distance ratio $r_d$ in peak time and nonpeak time for all OD pairs. The time ratio of a single trip, $r_t$, is determined as the actual travel time of the trip divided by the average travel time. Also, the time ratio $r_t$ in peak time and nonpeak time are derived for all OD pairs. As seen in SI, Figs. S2 and S3, the distributions of distance ratios and time ratios during peak and nonpeak hours have tent shapes, with the highest probability density when the ratio is about 1 and decreasing when the ratio is smaller or larger than 1. The distribution of these two ratios can be divided into two segments at the ratio of 1. The part with ratios smaller than 1 has a narrow distribution and lacks regularity, while the part with ratios larger than 1 has a wider distribution and is more meaningful in terms of actual traffic. In this paper, we mostly focus on the ratios larger than 1 (i.e., $r_d > 1$ and $r_t > 1$), which means that single trip distances and times are longer than the average distance and average time.

After extracting the distance ratio and time ratio for all OD trips, we explore their distribution function forms by using the Akaike weights (see Methods). The results of Beijing and New York are shown in Fig. 2, and the results of Chengdu, Shenzhen, and Chicago are shown in SI, Fig. S5. Interestingly, the distributions of distance ratios $r_d$ during peak ($P_p(r_d)$) and nonpeak ($P_n(r_d)$) periods of these five cities are best fitted by a power-law scaling,

$$P_p(r_d) \sim (r_d)^{-\alpha_1}, \tag{1}$$
$$P_n(r_d) \sim (r_d)^{-\alpha_2}. \tag{2}$$

Here, $\alpha_1$ and $\alpha_2$ are the power-law exponents during peak and nonpeak hours, respectively. The power-law distributions of distance ratios suggest that, though many travel distances are close to the average distance, a non-negligible number of larger scales of travel distances also exist in each city, including considerably long travel distance compared to the mean distance. On one hand, with the frequent occurrence of traffic congestion in urban areas, taxi drivers may take long detours to avoid the congested roads. On the other hand, drivers may also take large distances for the purpose of increasing revenue. These behaviors may be the origin of the power-law distribution of distance ratios. Such phenomenon also implies that the distance heterogeneity of taxi trips can be described by a single power-law function.

We also find that the exponent $\alpha_1$ in peak hours is larger than the exponent $\alpha_2$ in nonpeak hours in all five cities (see Table 1 and SI, Table S2). This finding indicates that all cities are likely to have less large travel distances in peak hours compared to nonpeak hours. A possible reason for this will be discussed later. Further, we use the Kolmogorov-Smirnov (KS) test (see Methods) to examine whether the exponents of distance ratios of each day have a similar behavior. Since the data for a single day is limited and the results are heavily influenced by noise, we combine two adjacent days to get reasonable statistical results. We show the results of Beijing and New York in Fig. 3A and C, and observe that the distributions of peak exponents and nonpeak exponents for these two cities are relatively narrow and significantly distinguishable from each other. Moreover, the peak exponents are usually larger than the nonpeak exponents. Similar results are also found in other cities (SI, Fig. S6). It is plausible that the different scaling laws of the distance ratio found in different travel periods of all cities reflect the different structural characteristics of travel in different cities.

Next, we analyze the scaling properties of time ratios $r_t$. We show the distributions for Beijing and New York in Fig. 2, and for the other three cities in SI, Fig. S5. We find that the Akaike weights (see Methods) favor a power-law distribution for both time ratios $r_t$ in both, peak hours ($P_p(r_t)$) and nonpeak hours ($P_n(r_t)$),

$$P_p(r_t) \sim (r_t)^{-\alpha_3}, \quad (3)$$
$$P_n(r_t) \sim (r_t)^{-\alpha_4}, \quad (4)$$

with the exponent $\alpha_3$ for peak hours and the exponent $\alpha_4$ for nonpeak hours. The power-law distributions reveal that there is a broad range of time ratios, including cases where single trip times are much longer than the average. This scaling law can help us to understand the diversity of taxi travel time, estimate the quality of taxi routes and evaluate the traffic conditions of cities.

The power-law exponents of time ratios exhibit a similar pattern to those of distance ratios: the exponent $\alpha_3$ in peak period is larger than the exponent $\alpha_4$ in the nonpeak period (see Table 1 and SI, Table S2). The results suggest that it is less likely to have large travel times in peak periods compared to nonpeak periods. Also, we use KS test (see Methods) to compare the distributions of peak exponents and nonpeak exponents of time ratios per day. The results demonstrate that the peak exponents are generally larger than the nonpeak exponents (Fig. 3 and SI, Fig. S6). The time ratios in peak and nonpeak periods follow a different power law, which implies different traffic properties in the two periods and different strategies should be adopted for traffic management.

The following question can be naturally raised: though the network structural topology of the same city is the same, why do the distance ratio and the time ratio behave differently in the different travel periods? More specifically, why are the exponents systematically larger in peak hours, i.e., shorter detour trips? A plausible explanation is as follows. The travel demand in peak hours is far greater than that in nonpeak hours, yielding massive available passengers to drivers. Thus, drivers probably prefer not to travel long distances and times during rush hours because they can easily find new passengers and increase their revenue. To test our hypothesis, we calculate the waiting time $\tau$ of taxis, i.e., the time interval between two adjacent occupied trips, in peak hours and nonpeak hours. As seen in SI, Fig. S7, the mean value of the waiting time during nonpeak hours in all five cities is typically twice as long as that during peak hours. Moreover, we also examine the average waiting time $\tau_m$ in peak and nonpeak periods for each day (SI, Fig. S8), and observe that the average waiting time in peak periods is always significantly less than that in nonpeak periods. The mean and standard deviation of the average waiting time $\tau_m$ are summarized in Table 1 and SI, Table S2. Thus, our results support the hypothesis that since drivers can easily find passengers they are motivated to shorten their trips during peak periods, so that large-scale travel distances and travel times are less likely to occur during this period.

Furthermore, we also notice that the power-law exponents of the distance ratio $r_d$ of Shenzhen and New York are significantly larger (i.e., shorter distances) than those of the other cities (see Table 1 and SI, Table S2), which arises the question of the origin to this phenomenon. We hypothesize that it might be related to the efficiency of the road network structure of the city. To this end, we examine this issue from the perspective of road network density [39] (SI, Fig. S9). It is reasonable to assume that higher density of major streets leads to more efficient traffic. To test this, we calculate in all five cities, the road densities separately for all five types of major roads (All), including motorway, trunk, primary, secondary and tertiary, as well as for the first four types of major roads (Main). The road density distribution is shown in SI, Fig. S10, noting that the mean values of All and Main road densities are larger in Shenzhen and New York than in the other cities, especially for the Main road density (summarized in Table 1 and SI, Table S2). Thus, a reasonable explanation for the large exponents of distance ratios is that the road conditions may be better in Shenzhen and New York (the Main road density is larger), making it easier for drivers to perform shorter detours (exponents of the distance ratio are larger).

Focusing on the distance ratio $r_d$ and time ratio $r_t$ of the five cities, especially New York, we find that these two ratios can be a large number (in Fig. 2 and SI, Fig. S5). Our tests suggest that the large values of the ratio ($r_d$ and $r_t$) during peak and nonpeak periods are dominated by relatively short Euclidean distances of OD pairs $d_{OD}$. As seen in SI, Figs. S11 and S12, the majority of trips with ratios larger than 10 occur between OD pairs with $d_{OD}$ shorter than 4 km. Moreover, we divide all trips into different groups (see SI, Table S3) according to the Euclidean distance of OD pairs $d_{OD}$, and explore the distribution of the distance ratio $r_d$ and time ratio $r_t$ during peak and nonpeak hours for each group. The results show that a large fraction of trips in each city (over 85% in four cities except Chicago which is 74%) occur between OD pairs with $d_{OD}$ less than 4 km, and the Akaike test indicates that the power-law mainly appears in the travel between these OD pairs (see SI, Table S3).

We have separately analyzed above the spatio-temporal scaling laws of taxi trips, but an important question is what is the relationship between times and distances, and which new insights can we learn from such a relation? In fact, the time required to travel a long taxi route may be long or short due to traffic conditions. It is critical to understand the correspondence relation between distances and times. For this, we first analyze the correlation between distances and times at the average level of trips between OD pairs, and then explore the correlation between distance ratios and time ratios at the single trip level. Fig. 4A and C show the relationship between average distance $d_m$ and average time $t_m$ for all OD pairs in Beijing and New York. We observe that average distances and average times are significantly and highly correlated by calculating the correlation $C$ and significance $W$ (Methods) (see SI, Fig. S13). Interestingly, we also find that the correlation $C$ is smaller in peak hours compared to nonpeak hours. Similar results are also found in other cities (SI, Fig. S14 A, C, and E). For a plausible reason see below. Further, to explore the changes in the correlation between distances and times in different distance ranges, we extract datasets with average distances of OD pairs larger than a given threshold (i.e., 0, 1, 2, 3, 4, 5, 6, 7 km) and then calculate the correlation between average distances and average times for each dataset. The results for Beijing and New York are shown in Fig. 4B and D, as the average distance threshold increases, the correlation decreases. Also note that the correlations are smaller during peak hours compared to nonpeak hours, which is consistent with the results for other cities (SI, Fig. S14 B, D, and F). Our findings suggest that (i) the longer the typical distance of OD pairs, the larger the fraction of taxis that do it in order to save time and (ii) more long-distance trips are made to save time during peak periods compared to nonpeak periods. As discussed above, the differences between peak and nonpeak hours may be caused by the availability of more passengers during peak hours, and saving time means higher revenue.

Next, we examine whether the correlation between distance ratios and time ratios of *individual* trips has a similar behavior. To this end, we use a similar way to extract the datasets with distance ratios of trips larger than a given threshold (of 1, 2, 3, 4, 5, 6, 7) and calculate the correlation between the distance ratio and time ratio for each dataset. As seen in SI, Fig. S16, as the distance ratio threshold increases, the correlation of nonpeak periods slightly decreases, while the correlation of peak periods decreases significantly and is typically smaller than the nonpeak correlation. Our results support the above hypothesis from a different perspective: trips that take long distances to save time are more likely to occur during peak periods, in particular for trips with large distance ratios, than during nonpeak periods.

To further test and support our hypothesis, we calculate directly the fraction of drivers who choose long distances to save time during peak and nonpeak hours, which we call saving drivers. We extract datasets with distance ratios larger than a given threshold (from 1 to 3 with an interval of 0.2) and calculate the fraction of drivers with time ratios smaller than 1 in each dataset. To eliminate noise, only drivers that do it for more than 50% of their trips are considered. That is, we consider a driver to be a true saving driver only if most of his/her long distance trips have time ratios smaller than 1. We find indeed, that the fraction of saving drivers during peak hours is higher than that during nonpeak hours in all five cities, with New York having the highest fraction at 20% in peak hours and 10% in nonpeak hours, while Beijing, another capital city, has a relatively low fraction at 11% in peak hours and 7% in nonpeak hours when the distance ratio threshold is 1.2 (see SI, Fig. S17). Our results reveal that, during peak hours, more drivers try to shorten the time of their trips by choosing long distances to save time. This can explain the lower correlation between distance ratios and time ratios during peak hours compared to nonpeak hours.

To further explore the possible origin of the power-law distribution of the distance ratio and time ratio for taxi drivers, we analyze a commonly used traffic assignment model, the stochastic user equilibrium (SUE) model [40] (see SI, Note 1). In this model, we use as input the trips in each of the OD pairs, and then analyze and test the distribution of these two ratios. Actually, the travel cost of each route known by drivers is only an estimate of the actual cost. The SUE model assumes that drivers choose the route with the minimum perceived (estimated) cost. In the equilibrium system state, no driver can reduce his perceived cost by unilaterally changing paths. Therefore, the SUE model is an indeterminate traffic assignment method, and multiple paths are chosen with different probabilities during path selection. Based on the SUE model, we assign a given travel demand $D$ to each path in a small-scale network, the well-known Sioux Falls network (in SI, Fig. S18), which is commonly used for numerical studies of traffic assignment [41–43]. The path-based Method of Successive Averages (MSA) is used to solve our SUE problem (see SI, Note 2). The path choice set for each OD pair is generated before the assignment, using the $k$-shortest path method [44] and setting $k$ to 7. Also, we assume that the dispersion parameter $\theta$, a measure of drivers' perception of travel costs, is equal to 1 based on empirical evidence [45].

After the traffic assignment, we obtained the trips of each path and calculated distance ratios and time ratios for each OD pair. Fig. 5A shows the distribution of the distance ratio and time ratio for the given original demand, which is found to follow a power law. Moreover, we find that increasing the original demand to a certain level, such as four times the original demand (Fig. 5B), the power-law exponent of these two ratios also increases. Thus, the model results are consistent with the findings from the actual data: when travel demand highly increases, like during peak hours, traffic can be very congested, resulting in higher perceived travel costs for drivers, and thus they are less likely to take large distances and large times. The model suggests that the power-law distributions of the distance and time ratios are the result of drivers' individual choice behavior and are influenced by random utility (i.e., drivers' perceived travel costs). Furthermore, sensitivity analysis of the travel demand $D$, the dispersion parameter $\theta$, and the size of the path choice set $k$ are performed to understand the impact of these factors on the distribution of these two ratios (shown in SI, Fig. S19).

**Conclusions**

Human mobility within cities is strongly correlated with urban traffic. Increased travel exacerbates traffic congestion, and traffic congestion, in turn, influences people's choice of travel routes. Earlier studies have shown that human movement has very broad scales represented by long-tail power-law distributions of traveling times and distances [13, 33]. However, the extent of spatio-temporal variation in people's travel routes for the *same* OD pair, which is important for mitigating traffic problems, has not been studied so far to the best of our knowledge. Based on taxi data of five metropolises in two countries, China and USA, we explored the scaling and universality features of the variability of intra-city human travel routes. Considering the significant difference in traffic conditions during peak and nonpeak hours, these two periods are analyzed separately. We examine the distance ratio and time ratio to measure the scale of spatio-temporal deviation of actual travel paths from the average (typical) travel path. We find that both ratios follow long-tail power-law distributions. This result suggests that a significant fraction of travel routes are much longer than the average route (see SI, Table S4). Surprisingly, we also find that the power-law exponent is larger during peak hours than during nonpeak hours in all analyzed cities. Our results suggest that shorter travel distances and times in the peak period are due to the availability of more passengers represented by the lower average waiting times in this period, so that drivers are motivated to shorten their trips and take another passenger. Therefore, with the aid of traffic management measures, such as staggered travel, it could be possible to change drivers' route selection decisions by adjusting the taxi demand. We also conclude that the power-law exponents of the distance ratio in different cities are affected by the urban road network structure, and some cities are significantly less likely to generate long distances, possibly due to their high density of major roads. Thus, increasing the density of efficient roads could provide a tool for reducing long detours. Moreover, we find a high correlation between distances and times, and the correlation is smaller in peak hours than in nonpeak hours. This result could be understood by the fact that during peak hours, due to the availability of many passengers, more drivers try to shorten the time by choosing long distances and thus increase their revenue. Finally, we apply an indeterminate traffic assignment model [37] to try to understand the origin of the scaling power-laws for the distributions of the distance and time ratios. The model results demonstrate that the power-law scaling of taxi routes is indeed the outcome of drivers' individual choices and is influenced by random utility, which provides insight into transportation economic modeling. The present study can help to assess urban traffic conditions and provide guidance for urban traffic management, and can also be used to evaluate the money-loss of passengers based on the fraction of travel with very large distance ratios and time ratios.

**Methods**

**Taxi dataset.** To have a reliable measurement, we exclude trips in which both O and D are in the same zone and only study OD pairs with over 100 trips.

**Akaike weights.** Using this method, we test whether the given dataset $x = \{x_1, x_2, x_3, \cdots, x_n\}$ fits better with a power-law tail or an exponential tail [46–48]. Their probability density functions $p(x) = Cf(x)$, consisting of the basic function form $f(x)$ and the appropriate normalized constant $C$, are shown below. Considering the tail to start at $x_{\min}$, the probability density function of the power-law distribution is defined as:

$$p(x) = (\alpha - 1)x_{\min}^{\alpha-1} x^{-\alpha}, \quad (5)$$

where $(\alpha - 1)x_{\min}^{\alpha-1}$ is the normalized constant and $\alpha$ is the power-law exponent. The probability density function of the exponential distribution is defined as:

$$p(x) = \lambda e^{\lambda x_{\min}} e^{-\lambda x}, \quad (6)$$

where $\lambda e^{\lambda x_{\min}}$ is the normalized constant and $\lambda$ is the exponential rate parameter.

The fitting parameters are computed by the maximum likelihood estimation (MLE) [46,47]. The Akaike information criterion (AIC) [48] is employed to choose the best-fitted distribution. For the candidate model $i$ ($i = \{1, 2\}$), the corresponding $AIC$ score is computed by $AIC_i = -2\log L_i + 2K_i$, where $L_i$ is the likelihood function and $K_i$ is the number of parameters in the model $i$.

The Akaike weights can be considered as relative likelihoods being the best model for the observed data. Let

$$AIC_{\min} = \min\{AIC_i\}, \quad (7)$$

$$\Delta_i = AIC_i - AIC_{\min}. \quad (8)$$

Then the Akaike weight $W_i$ is calculated by

$$W_i = \frac{e^{-\Delta_i/2}}{e^{-\Delta_1/2} + e^{-\Delta_2/2}}. \quad (9)$$

An Akaike weight is a normalized distribution selection criterion [49]. Its value is between 0 and 1. The larger the value is, the better the distribution is fitted.

**Calculation of correlation and significance.** To measure the correlation between distance and time, we calculate the Pearson correlation coefficient and the significance indicator between the two variables. Pearson coefficient can reflect the degree of linear correlation between two variables. For two random variables $X = \{x_1, x_2, \cdots, x_n\}$ and $Y = \{y_1, y_2, \cdots, y_n\}$, the correlation coefficient $C_{X,Y}$ is

$$C_{X,Y} = \frac{\operatorname{cov}(X,Y)}{\sigma_X \sigma_Y} = \frac{E\left[(X-\mu_X)(Y-\mu_Y)\right]}{\sigma_X \sigma_Y}, \tag{10}$$

where $\mu_X$ and $\mu_Y$ are the mean values of $X$ and $Y$, and $\sigma_X$ and $\sigma_Y$ are the standard deviations of $X$ and $Y$.

To determine whether the correlation between $X$ and $Y$ is significant, we shuffle the data series of the variable $Y$, and then calculate the cross-correlation coefficients $C_{X,Y(k)}$ of variables $X$ and $Y(k)$ as follows,

$$C_{X,Y(k)} = \frac{\operatorname{cov}(X,Y(k))}{\sigma_X \sigma_Y} = \frac{\sum_{i=1}^{n}(x_i - \bar{X})(y_{i+k} - \bar{Y})}{\sqrt{\sum_{i=1}^{n}(x_i - \bar{X})^2}\sqrt{\sum_{i=1}^{n}(y_{i+k} - \bar{Y})^2}}, \tag{11}$$

where $Y(k)$ indicates that the data series of $Y$ is circularly shifted to $k$ data points, $k = 0, \cdots, n-1$. If $k = 0$, $C_{X,Y(k)} = C_{X,Y}$.

After determining the cross-correlation coefficients $C_{X,Y(k)}$ of each $Y(k)$, the significance indicator $W_{X,Y}$ of variables $X$ and $Y$ can be given

$$W_{X,Y} = \frac{C_{X,Y(0)} - \operatorname{mean}\left(C_{X,Y(k^*)}\right)}{\operatorname{std}\left(C_{X,Y(k^*)}\right)}, k^* = 1, \ldots, n-1. \tag{12}$$

**Two-sample Kolmogorov–Smirnov test.** We use the Kolmogorov-Smirnov test (KS test) to compare two sample distributions (two-sample KS test). The KS statistic is:

$$D_{n,m} = \sup_x | F_{1,n}(x) - F_{2,m}(x) |, \tag{13}$$

where $F_{1,n}$ and $F_{2,m}$ are the empirical distribution functions of the first and second samples, respectively, and sup is the supremum function.

The functions $F_{1,n}$ and $F_{2,m}$ are defined as:

$$F_{1,n}(x) = \frac{1}{n}\sum_{i=1}^{n} I_{[-\infty,x]}(X_i), \tag{14}$$

$$F_{2,m}(x) = \frac{1}{m}\sum_{i=1}^{m} I_{[-\infty,x]}(X_i), \tag{15}$$

where $n$ and $m$ are the sizes of the first and second samples, respectively. $I_{[-\infty,x]}(X_i)$ is the indicator function, which is equal to 1 if the observation $X_i < x$ and equal to 0 otherwise.

For large samples, the null hypothesis is rejected at significance level $\alpha$ if

$$D_{n,m} > c(\alpha)\sqrt{\frac{n+m}{n \cdot m}}, \qquad (16)$$

where the $c(\alpha)$ value can be obtained by $c(\alpha) = \sqrt{-\ln\left(\frac{\alpha}{2}\right) * \frac{1}{2}}$ [50]. We set the significance level $\alpha$ to 0.05, and the corresponding $c(\alpha)$ is 1.358.

**Traffic assignment.** Traffic assignment is a mature field that aims to integrate travel demand with road infrastructure, to better understand traffic, and has been extensively studied by urban and transportation planners. In this work, we use a stochastic user equilibrium model for traffic assignment [37] (see SI Note 1). A static, path-based assignment algorithm is then employed for the solution (see SI Note 2).

ACKNOWLEDGMENTS. H.S. and J.W. acknowledge support from the National Natural Science Foundation of China (Grants 91846202 and 71890972/71890970). J.W. also acknowledges support from the State Key Laboratory of Rail Traffic Control and Safety (RCS2020ZZ001). S.H. thanks the Israel Science Foundation, the Binational Israel-China Science Foundation (Grant No. 3132/19), the BIU Center for Research in Applied Cryptography and Cyber Security, NSF-BSF (Grant No. 2019740), the EU H2020 project RISE (Project No. 821115), the EU H2020 DIT4TRAM, and DTRA (Grant No. HDTRA-1- 19-1-0016) for financial support.

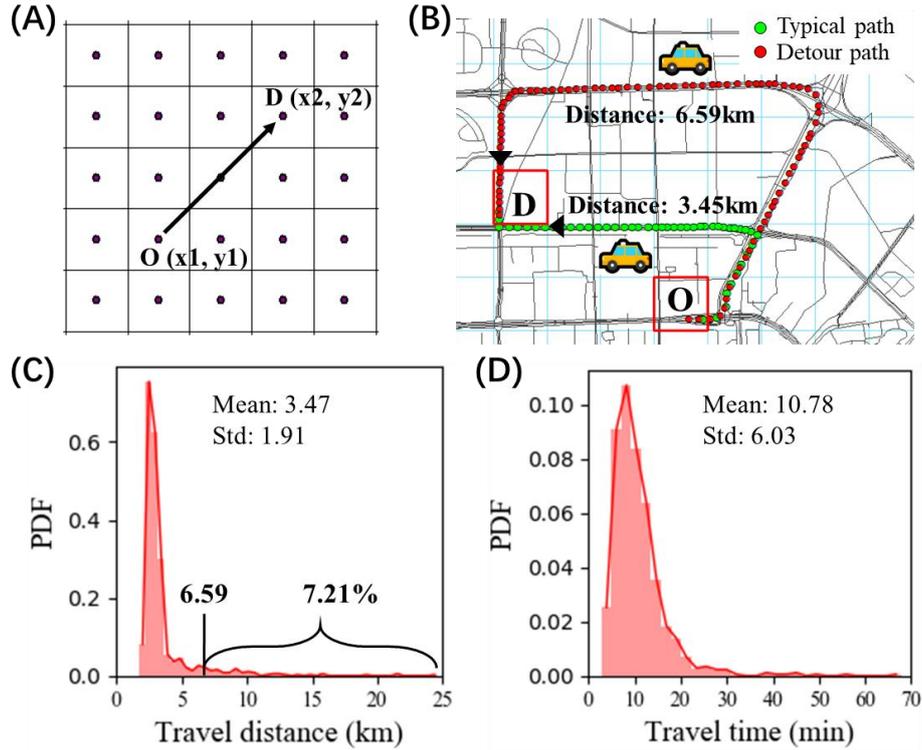

**Fig. 1. Taxi trips illustration. (A) Grid representation of a city.** We divide the city area into a grid of sizes 0.5x0.5 km. The Euclidean distance of the OD pair is defined as the distance between the grid centers $d_{OD} = \sqrt{(x_1 - x_2)^2 + (y_1 - y_2)^2}$. **(B) Taxi trips paths.** Illustration of an OD pair (red squares) in Beijing. Many possible paths are connecting this OD pair, and most trips follow the typical path (green circles) with a length close to the average distance, but there are some exceptions. Trips follow a detour path (red circles), which might be in order to avoid congestion, gain more benefits, or due to visiting multiple destinations, and are characterized by a longer distance than the average distance. The length of the detour path here is about twice as long as the typical path. **(C) The travel distance and (D) travel time distributions of this OD pair.** One can see that 7.21% of trips are even longer than the detour path in (B), and some are even seven times longer than the average distance (3.47 km). Likewise, travel times vary significantly, with some more than six times longer than the average time (10.78 min).

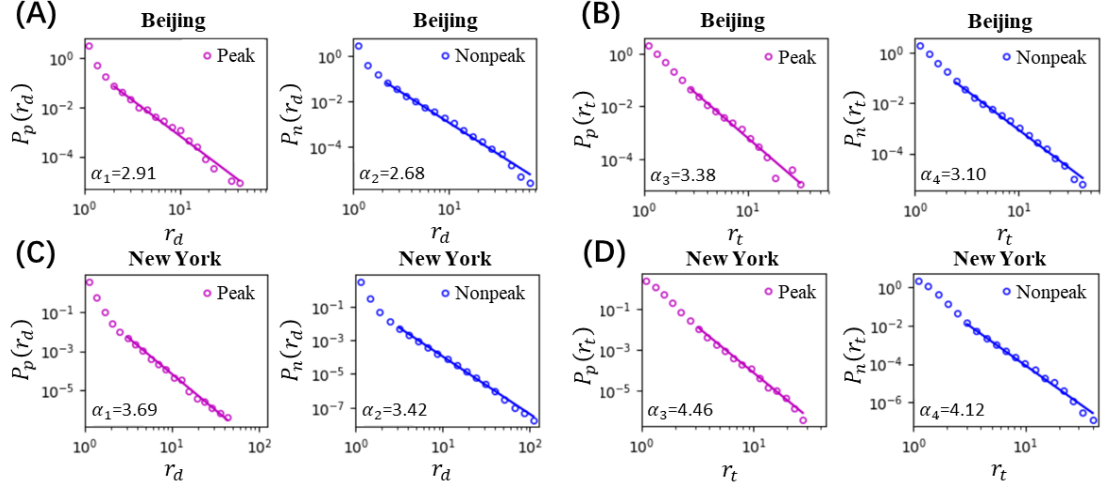

**Fig. 2. Probability distributions of distance ratios and time ratios of taxi trips.** The distributions of the distance ratio for all OD pairs in (A) Beijing and (C) New York follow a power law for both peak and nonpeak hours. The lines represent the best fit above a threshold evaluated in SI, Fig. S4, and the exponents are obtained using the maximum likelihood estimation based on these thresholds. Similar power-law distributions are found for the time ratios in (B) Beijing and (D) New York. In both the distance ratio and time ratio distributions, the exponent of peak hours (i.e., 7:00-9:00 and 17:00-19:00) is larger than that of nonpeak hours. Other cities show similar results and can be found in the SI, Fig. S5.

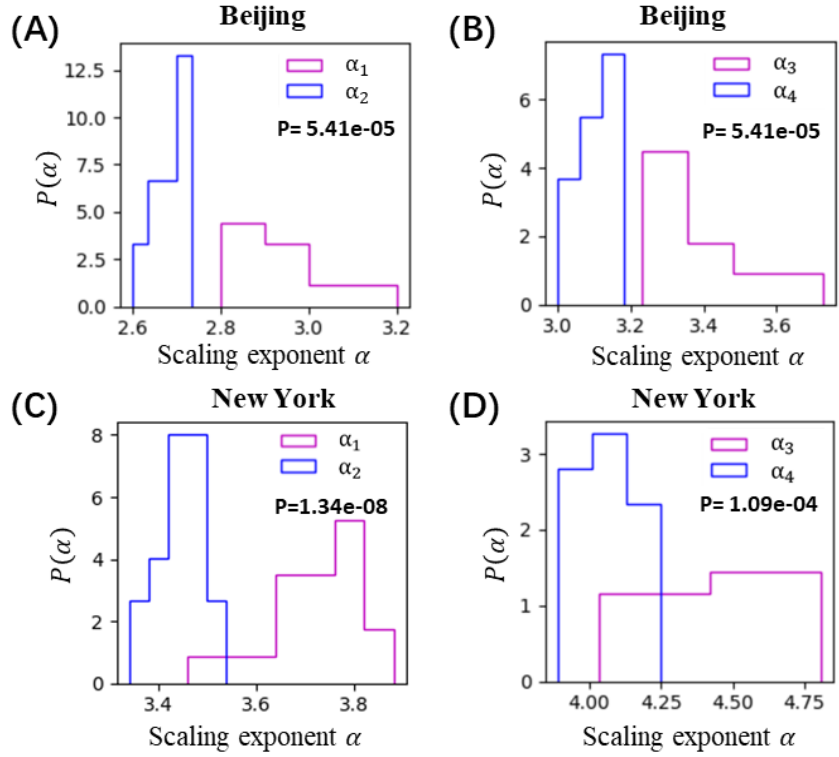

**Fig. 3. The distribution of scaling exponents of the distance ratio and time ratio on different days.** The distribution of power-law exponents of the distance ratio during peak and nonpeak hours on pairs of consecutive days in (A) Beijing and (C) New York. Both cities show power-law distributions for all days. Moreover, the KS test of the exponent distribution in peak and nonpeak hours shows that the p-value (P) is less than 0.05, revealing that the distributions of the peak and nonpeak exponents are significantly different from each other. Similar behavior is also found for the power-law exponent distribution of the time ratio on pairs of consecutive days in (B) Beijing and (D) New York. For the exponent distribution of distance ratios and time ratios, the peak exponents are mostly larger than the nonpeak exponents. Other cities show similar results and can be found in the SI, Fig. S6.

|  | Beijing | | New York | |
| --- | --- | --- | --- | --- |
|  | Peak (*p*) | Nonpeak (*n*) | Peak (*p*) | Nonpeak (*n*) |
| $r_d$ exponent | 2.91 | 2.68 | 3.69 | 3.42 |
| $r_t$ exponent | 3.38 | 3.10 | 4.47 | 4.12 |
| $\tau_m$ (min) | 10.59±0.54 | 20.55±1.00 | 9.07±1.15 | 19.10±1.56 |
| Main road density (km/km$^2$) | 4.15±3.68 | | 7.75±4.31 | |

**Table 1. Taxi routes parameters.** The power-law exponents of the distance ratio and the time ratio are given for peak and nonpeak hours in Beijing and New York. Note that they are found to be larger in peak hours than in nonpeak hours. Higher exponents indicate fewer longer trips than the typical path. We hypothesize that this might be related to the smaller average waiting time in peak hours compared to nonpeak hours. Thus, the driver has a motivation to shorten the trip and take a new passenger. Moreover, the exponents of the distance ratio of New York and Shenzhen (see SI, Table S2) are significantly larger than those of other cities, which might be related to their larger Main road density (see SI, Table S2), making it easier for drivers to take shorter detours. The parameters of other cities can be found in the SI, Table S2.

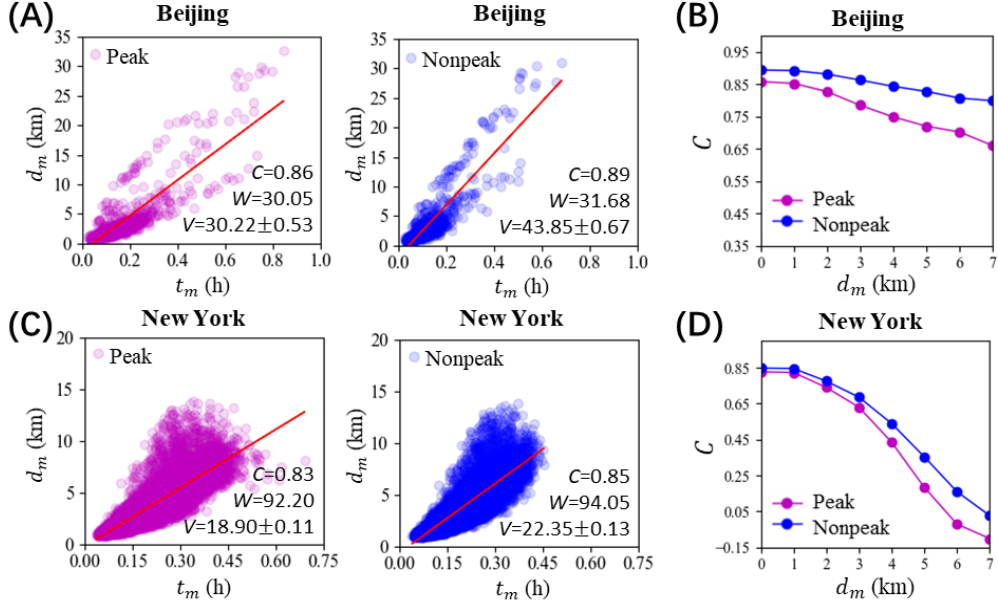

**Fig. 4. Average distance vs average time in peak and nonpeak hours for (A) Beijing and (C) New York.** Here, we show the scatter plots of the average distance vs. average time for all OD pairs. $C$ and $W$ represent the correlation coefficient and significance, respectively. As seen, there is a significantly high correlation between the average distances and the average times, and the correlation is larger in nonpeak periods compared to peak periods. The slope $V$ of the fitted line (red line) represents the average velocity (in km/h) calculated from the ratio between the average distance and average time. The average velocity is, as expected, smaller at peak compared to nonpeak due to traffic jams. **Correlation between average distance and average time during peak and nonpeak periods at different average distance ranges for (B) Beijing and (D) New York.** We extract datasets with average distance of OD pairs $d_m$ larger than a given threshold for peak and nonpeak periods separately, and then calculate the correlation between the average distances and average times for all OD pairs in each dataset. We set eight thresholds, i.e., 0, 1, 2, 3, 4, 5, 6, 7 km. Other cities show similar results and can be found in the SI, Fig. S14.

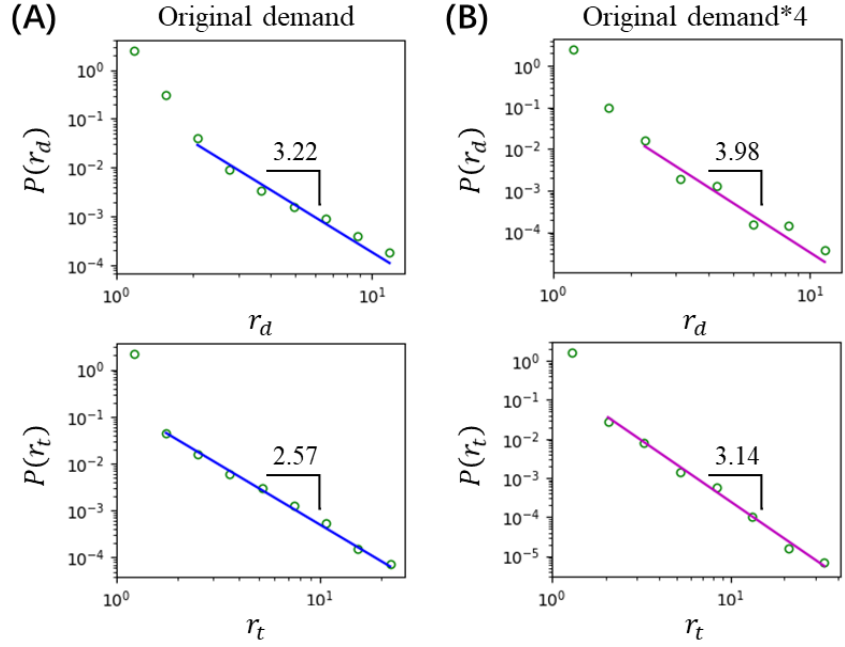

**Fig. 5. Distance ratio and time ratio distributions based on the (A) original demand and (B) four times the original demand using the traffic assignment model.** Using the Akaike weights (see Methods), we find that the distance ratio and time ratio distributions for all OD pairs follow a power law. The lines represent the best fit above the calculated threshold (2 for distance ratios and 1.5 for time ratios) and the exponents are obtained using the maximum likelihood estimation based on their thresholds. A similar behavior is found for the distributions of the two ratios when increasing the demand to four times its original size. Moreover, the power-law exponent of the increased demand is found to be larger than that of the original demand, which is in agreement with the findings in the real data.